\documentclass[aps,prd,amsfonts,amssymb,amsmath,showpacs]{revtex4}

\usepackage{epsfig}
\usepackage{bm}

\begin{document}

\title{Conformally coupled scalar solitons and black holes
with negative cosmological constant}

\author{Eugen Radu}
\email{radu@thphys.nuim.ie}
\affiliation{Department of Mathematical Physics, National University of Ireland,
Maynooth, Ireland.}
\author{Elizabeth Winstanley}
\email{E.Winstanley@sheffield.ac.uk}
\affiliation{Department of Applied Mathematics, The University of Sheffield,
Hicks Building,
Hounsfield Road, Sheffield, S3 7RH, United Kingdom.}

\date{\today}

\begin{abstract}
We present arguments for the existence of both globally regular and black hole
solutions of the Einstein equations with a conformally coupled scalar field,
in the presence of a negative cosmological constant, for
space-time dimensions greater than or equal to four.
These configurations approach asymptotically anti-de Sitter spacetime and
are indexed by the central value of the scalar field.
We also study the stability of these solutions, and show that, at least
for all the solutions studied numerically, they are linearly stable.
\end{abstract}

\pacs{04.20.Jb, 04.40-.b, 04.70.Bw}

\maketitle

%%%%%%%%%%%%%%%%%%%%%%%%%%%%%%%%%%%%%%%%%%%%%%%%%%%%%%%%%%%%%%%%%%%%%%%%%%%%%%
%%%%%%                     Introduction
%%%%%%%%%%%%%%%%%%%%%%%%%%%%%%%%%%%%%%%%%%%%%%%%%%%%%%%%%%%%%%%%%%%%%%%%%%%%%%
\section{Introduction}
Within the various theories
violating the no-hair conjecture, the case of a conformally coupled
scalar field is of particular interest.
In asymptotically flat space
with no scalar self-interaction potential, this theory admits an
exact, closed form black hole solution, which
has a scalar field diverging on the event horizon \cite{Bekenstein:1974sf}.
However, it is known that this solution is unstable \cite{Bronnikov:1978mx}, and cannot
be considered as a valid example of a black hole with scalar hair.
There are also a number of theoretical results as well as numerical
evidence against the existence of black holes with scalar field hair
(with various couplings to the Ricci scalar curvature) in asymptotically
flat spacetime (see, for example, \cite{Winstanley:2005fu} for a recent discussion).

In an unexpected development, hairy black hole solutions
have been found in both theories with minimally as well as nonminimally
coupled scalar fields by considering asymptotically anti-de Sitter (AAdS)
boundary conditions \cite{Torii:2001pg,Winstanley:2002jt}.
Moreover, some of these solutions are found to be stable.
Exact four-dimensional black hole solutions of gravity
with a minimally coupled self-interacting
scalar field have been presented by
 Martinez, Troncoso and Zanelli (MTZ) \cite{Martinez:2004nb}
 and Zloshchastiev \cite{Zloshchastiev:2004ny}.

However, in many other theories admitting hairy black hole solutions,
these configurations survive in the limit of zero event horizon
radius, yielding particle-like, globally regular configurations.
 Motivated by this observation, we consider in this paper the case of
a conformally coupled scalar field in an $n-$dimensional
AAdS spacetime (with $n\geq 4$)
and look for both globally regular and black hole solutions.
In the black hole case, we extend the results of Ref. \cite{Winstanley:2002jt}
by considering  higher dimensional configurations.
Since a negative cosmological constant allows for the existence
of black holes whose horizon has nontrivial topology,
we consider,  apart from spherically symmetric solutions,
topological black holes also.
We find that the spherically symmetric solutions admit a nontrivial
regular limit, representing gravitating scalar solitons.
These configurations are indexed by the central value of the scalar field
and are found to be stable against linear fluctuations.

The outline of this paper is as follows: in Section
\ref{sec:model} we introduce
our model, the numerical results being presented in Section \ref{sec:num}.
The stability of our solutions is addressed in Section \ref{sec:stab}. Our
conclusions are presented in Section \ref{sec:conc}.

%%%%%%%%%%%%%%%%%%%%%%%%%%%%%%%%%%%%%%%%%%%%%%%%%%%%%%%%%%%%%%%%%%%%%%%%%%%%%%
%%%%%%                      The ansatz
%%%%%%%%%%%%%%%%%%%%%%%%%%%%%%%%%%%%%%%%%%%%%%%%%%%%%%%%%%%%%%%%%%%%%%%%%%%%%%
\section{The model}
\label{sec:model}
\subsection{The ansatz and field equations}
We consider the following action, which describes a self-interacting
scalar field $\phi $ with non-minimal coupling to gravity in $n-$dimensions
(throughout this paper we will use units in which $c=8\pi G=1$)
\begin{eqnarray}
\label{action}
S=\int d^{n}x \, {\sqrt {-g}} \left[
\frac {1}{2}\left( R -2\Lambda \right)
-\frac {1}{2} \left( \nabla \phi \right) ^{2}
-\frac {1}{2} \xi R \phi ^{2} -V(\phi ) \right] ,
\end{eqnarray}
where $R$ is the Ricci scalar curvature, $\Lambda=-(n-2)(n-1)/2\ell^2$ is the cosmological
constant and $\xi$ is the coupling constant.
For a minimally coupled scalar field $\xi=0$ and for conformal
coupling (which is the focus of this paper) $\xi=\xi_c=(n-2)/4(n-1)$.

The field equations are obtained by varying the action (\ref{action})  with respect
to field variables $g_{\mu \nu}$ and $\phi$
\begin{eqnarray}
\label{einst}
 \left( 1- \xi \phi ^{2} \right) G_{\mu \nu }
+g_{\mu \nu } \Lambda
&=&
\left( 1 - 2\xi \right) \nabla _{\mu } \phi \nabla _{\nu }
\phi
+\left(2 \xi -\frac {1}{2} \right) g_{\mu \nu } \left(
\nabla \phi \right)^{2}
\nonumber
\\
&&-2\xi \phi \nabla _{\mu } \nabla _{\nu } \phi
+2\xi g_{\mu \nu } \phi \nabla ^{2}  \phi
- g_{\mu \nu } V(\phi ),
\\
\label{KG}
\nabla^2 \phi - \xi  R\phi -
\frac {dV}{d \phi } &=&0.
\end{eqnarray}
The Ricci scalar expression which follows from these equations is
\begin{eqnarray}
R=-\frac{ 2(n-1)(\xi-\xi_c)(\nabla \phi)^2
+2\xi(n-1)\phi\frac{dV}{d\phi}-n(V(\phi)+\Lambda) }
{n/2-1+2(n-1)\xi(\xi-\xi_c) \phi^2 }.
\end{eqnarray}
Since for a negative cosmological constant
topological black holes may appear
(with a nonspherical topology of the event horizon),
we consider a general metric ansatz
\begin{eqnarray}
\label{metric}
ds^{2} =
\frac{dr^{2}}{H(r)}+ r^{2} d \sigma_{d-2,k}^2- H(r)e^{2\delta (r)} dt^{2},
\end{eqnarray}
%%%%%%%%%%%%%%%%%  two-surface of constant curvature %%%%%%%%%%%%%%%%%
where $d\sigma^2_{n-2,k}=d\psi^{2}+f^{2}_k(\psi) d\Omega_2^2$ denotes the line
element of an $(n-2)$-dimensional space $\Sigma_{k}$ with
constant curvature.
The discrete parameter $k$ takes the values $1, 0$ and $-1$
and implies the following form of the function $f_k(\psi)$:
\begin{equation}
f_k(\psi)=\left \{
\begin{array}{ll}
\sin\psi, & {\rm for}\ \ k=1 \\
\psi , & {\rm for}\ \ k=0 \\
\sinh \psi, & {\rm for}\ \ k=-1.
\end{array} \right.
\end{equation}
When $k=1$, the hypersurface $\Sigma_{1}$ represents a sphere, for $k=-1$,
it is a negative constant curvature space and it could be a
closed hypersurface with arbitrarily high genus under appropriate identifications.
For $k=0$, the hypersurface $\Sigma_{0}$ is a $(n-2)$-dimensional Euclidean space
(see $e.g.$ the discussion in \cite{Mann:1997iz}).

A convenient parametrization of the metric function $H$ is
\begin{eqnarray}
\label{H}
H(r)=k-\frac{2m(r)}{r^{n-3}}+\frac{ r^2}{\ell^2}.
\end{eqnarray}
Although a rigorous computation of the solutions' mass and action is a nontrivial
task for $\xi \neq 0$, we assume that
 the asymptotic value of $m(r)$ corresponds to the mass of our configurations,
up to some $n-$dependent factor.

Within this ansatz, we find the resulting equations
(where a prime denotes the derivative with respect to $r$)
\begin{eqnarray}
\nonumber
\label{eqs}
\frac{n-2}{2r}(1-\xi \phi^2)\left[H' -\frac{n-3}{r}(k-H)\right] -
\left(2\xi-\frac{1}{2}\right)H\phi'^2+\xi \phi \phi'(H'+2H \delta')
-2\xi \phi\nabla^2 \phi +V(\phi)+\Lambda=0,
\\ \nonumber
\frac{n-2}{r}(1-\xi\phi^2)\delta'-(1-2\xi)\phi'^2-2\xi\phi(\delta'\phi'-\phi'')=0,
\\
H \phi''+\phi'\left(H\delta'+H'+H\frac{n-2}{r}\right)-\xi R \phi -\frac{dV(\phi)}{d \phi}=0.
\end{eqnarray}
%%%%%%%%%%%%%%%%%%%%%%%%%%%%%%%%%%%%%%%%%%%%%%%
\subsection{Boundary conditions}
The asymptotic solutions to these equations can be systematically
constructed in both regions, near the origin (or event horizon)
and for $ r \gg 1$.

The corresponding expansion as $r \to 0$ is
(globally regular solutions may exist for $k=1$ only):
\begin{eqnarray}
\label{origin}
\nonumber
H(r) & = &1+\frac{2(-2\xi^2\phi_0^2R_0+\Lambda+V_0-2\xi\phi_0
V_0')}{(n-1)(n-2)(\xi\phi_0^2-1)}r^2+O\left(r^4\right),
\\
\nonumber
 \delta(r)& = & \delta_0+\frac{\xi \phi_0(\xi \phi_0 R_0+V_0')}{(n-1)(n-2)(\xi
\phi_0^2-1)}r^2+O\left(r^4\right),
\\
\phi(r) & =& \phi_0+\frac{\xi \phi_0R_0+V'_0}{2(n-1)}r^2+O\left(r^4\right),
\end{eqnarray}
where
\begin{equation}
R_0=\frac{n(V_0+\Lambda)-2
\xi(n-1)\phi_0V_0'}{n/2-1+2(n-1)\xi (\xi-\xi_c)\phi_0^2}
\end{equation}
is the Ricci scalar evaluated at the origin and
$V_0=V(\phi_0),~V_0'=dV/d\phi|_{\phi=\phi_0}$.

For black hole configurations with a regular, nonextremal event horizon at $r=r_h$,
the expression  near the event horizon is
\begin{eqnarray}
\label{eh}
\nonumber
H(r)&=&H'(r_h)(r-r_h)+O\left(r-r_h\right)^2,
\\
\nonumber
 \delta(r) &  = & \delta_h+\delta'(r_h)(r-r_h)+O\left(r-r_h\right)^2,
\\
\phi(r)&=&\phi_h+\phi'(r_h)(r-r_h)+\phi_2(r-r_h)^2+O\left(r-r_h\right)^3,
\end{eqnarray}
where
\begin{eqnarray}
\nonumber
H'(r_h)&=&(n-3)\frac{k}{r_h}+\frac{2r_h\big(\Lambda
+V_h-\xi\phi_h(\xi \phi_h R_h+V'_h)\big)}{(n-2)(\xi \phi_h^2-1)},
\\
\nonumber
\phi'(r_h) & = & \frac{\xi R_{h} \phi_h+V'_h}{H'(r_h)},
\\
\nonumber
\phi_2&=&\frac{\phi' (r_h)}{2r_h}\left\{-1+\frac{1}{2H_h' r_h}
\left[r_h^2 (\xi R_h+V''_h)+2k(n-3)\right]\right\}
+ \frac{1}{(\xi \phi_h^2-1)}
\left\{ \frac{(2\xi-1)\phi'(r_h)^2 r_h V_h'}{2(n-2)H'(r_h)}
\right.
\\
\nonumber
&&
\left.
+\xi \phi_h \phi'(r_h)
\left[ 1-\frac{k(n-3)}{2r_hH'(r_h)}+\frac{r_h}{2(n-2)H'(r_h)}(3 \xi
R_h+V''_h)\right]\right\} ,
\\
\delta'(r_h)&=&\frac{r_h[(2\xi-1)\phi'(r_h)^2+4\xi\phi_h \phi_2]}{(n-2)(\xi
\phi_h^2-1)+2\xi\phi_h \phi'(r_h)r_h},
\end{eqnarray}
and
\begin{eqnarray}
R_h=\frac{n(V_h+\Lambda)-2\xi
(n-1)\phi_hV'_h}{n/2-1+2(n-1)\xi(\xi-\xi_c)\phi_h^2}
\end{eqnarray}
is the Ricci scalar evaluated at the event horizon; noting that
$V^{(k)}_h=V^{(k)}(\phi)\big|_{\phi=\phi_h}$.

%%%%%%%%%%%%%%%%%%%%% asymptotics %%%%%%%%%%%
To analyze the $ r \gg 1$ region, we assume that
the geometry approaches asymptotically the AdS spacetime and that
the function  $m(r)$ does not diverge in the same limit.
These assumptions imply that $\lim_{r\to\infty}\phi(r)~=~0$ and
\begin{eqnarray}
\phi=\frac{c_1}{r^{n/2-1}}+\frac{c_2}{r^{n/2}}
 -\frac{c_3}{r^{n/2+1}}+\dots,
~~
H(r)=k-\frac{2M}{r^{n-3}}-\frac{2\Lambda r^2}{(n-2)(n-1)}+\dots ,
~~
\delta(r)= \frac{\delta_2}{r^n}+\dots,
\end{eqnarray}
and also impose some constraints on the scalar potential
(for example, in $n=4$, the potential
should satisfy as $r\to \infty$ the condition
$V=V'=V''=V'''=0$).
In the above relations $c_1,~c_2$ and $M$ are real constants which fix the values of the
other coefficients in the asymptotic expansion.
In the simplest case of a vanishing self-interaction potential, we find
\begin{eqnarray}
c_3 & = & -\frac {c_1k(n-1)(n-2)^2(n-4)}{16 \Lambda},\nonumber \\
\delta_2 &= &-\frac {c_1^2(n-1)(n-2)^3(n-4)k+8c_2^2n\Lambda}{16 n(n-1)(n-2)\Lambda} .
\end{eqnarray}

%%%%%%%%%%%%%%%%%%%%%%%%%%%%%%%%%%%%%%%%%%%%%%%%%%%%%%%%%%%%%%%
\subsection{Conformal transformation}
The conformal transformation \cite{Bekenstein:1974sf,Maeda:1988ab}
\begin{equation}
\bar{g}_{\mu \nu}=\Omega^{2/(n-2)} g_{\mu \nu},
~~{\rm with}~~\Omega =1-\xi \phi^2
\label{Omega}
\end{equation}
maps the original system (\ref{action}) onto a much simpler one involving
just a minimally coupled scalar field, but with a more complicated potential
(this transformation is valid only for those solutions with a nonvanishing $\Omega$).
For a conformally coupled scalar field, the new action principle takes the form
\begin{eqnarray}
\label{action-c}
S=\int d^{n}x \, {\sqrt {-\bar{g}}} \left[
\frac {\bar{R}-2\Lambda }{2}
-\frac {1}{2} \left( {\bar {\nabla }} \Phi \right) ^{2}
- U(\Phi ) \right] ,
\end{eqnarray}
where a bar denotes quantities calculated using the transformed metric
$\bar{g}$ and we define a new scalar field $\Phi$ as (for $\xi = \xi _{c}$):
\begin{eqnarray}
\label{psi}
\Phi=\frac{1}{\sqrt{\xi_c}} \rm{arctanh} \sqrt{\xi_c} \phi,
\end{eqnarray}
with a nonvanishing potential
\begin{eqnarray}
\label{Vpsi}
U(\Phi) =[\Lambda +V(\phi)]
\left[ \cosh (\sqrt{\xi_c} \Phi ) \right] ^{\frac {2n}{n-2}}
-\Lambda .
\end{eqnarray}
The main advantage of the rescaled frame is that the field equations are
much simpler.
However, the potential (\ref{Vpsi}) is unphysical (for example,
with $V\equiv 0$, it is negative everywhere).

In the transformed frame, the metric ${\bar {g}}_{\mu \nu }$
takes the form (\ref{metric}), but with the quantities $H$ and $\delta $
replaced by ${\bar {H}}$ and ${\bar {\delta }}$ respectively.
In addition, there is a new radial co-ordinate ${\bar {r}}= \Omega ^{1/(n-2)} r$,
which is a good co-ordinate as long as
\begin{equation}
{\cal {A}} = (n-2)\Omega - 2 r \xi _{c} \phi \phi '>0.
\label{calA}
\end{equation}
This is an additional constraint on the scalar field $\phi  $ required for the
conformal transformation to be valid.
For all our numerical solutions, the conditions $\Omega,~{\cal {A}}>0$ are satisfied.
%%%%%%%%%%%%%%%%%%%%%%%%%%%%%%%%%%%%%%%%%%%%%%%%%%%%%%%%%%%%%%%
%% NUMERICAL SOLUTIONS
%%%%%%%%%%%%%%%%%%%%%%%%%%%%%%%%%%%%%%%%%%%%%%%%%%%%%%%%%%%%%%%
%
\section{Numerical results}
\label{sec:num}
As analytic solutions to the coupled nonlinear equations
(\ref{eqs})
appear to be intractable for every dimension,
except for the $n=4,~k=-1$
black hole solution found in \cite{Martinez:2004nb},
the resulting system has to be solved numerically.

In this section we discuss mainly the case of a
conformally coupled scalar field
without a self-interaction potential.
Since for $V(\phi)=0$ the field equations are invariant under the transformation
$\phi \to -\phi$, only positive values of $\phi_i$ are considered.
Here $\phi_i$ denotes the initial value of the scalar
field at the $r=r_0$ (with $r_0=(0,~r_h)$ for regular
and black hole solutions, respectively).
Also, by  rescaling the radial coordinate (together with $m(r)$),
we can set $\Lambda=-(n-1)(n-2)/2$ ($i.e.$ $\ell=1$)
without any loss of generality.

We follow the usual approach and, by using a
standard ordinary  differential
equation solver, we
evaluate  the  initial conditions at $r=r_0+10^{-6}$
for global tolerance $10^{-12}$,
and integrate  towards  $r\to\infty$.
In this way we find that nontrivial solutions
may exist in any dimension $n \geq 4$
(both black hole and regular solutions exist also in three spacetime
dimensions; however, their properties are somewhat
special and we do not discuss them here).
Black hole solutions seem to exist for any values of the parameters
$(k,~\phi_h,~r_h)$ satisfying $H'(r_h)>0$.

Typical profiles are presented in Figure 1 for regular configurations
and in Figures 2, 3 for  black hole solutions.
The dependence of the mass parameter $M$
 and $e^{2\delta(r_0)}$ on $\phi_i$ is plotted in Figure 4-6
(note the occurrence of negative values of $M$ for $k=-1$ black holes,
 a common situation in topological black hole physics).

The properties of the configurations can be summarized as follows:
\begin{enumerate}
\item
For $n\geq 4$, AAdS solutions exist for any values of
$\phi_i$ in the interval
$0<\phi_i<1/\sqrt{\xi_c}$;
\item
 The scalar field interpolates monotonically between $\phi_i$ and zero
and has no nodes;
\item
The value of the metric function $e^{2\delta}$ at the origin
(event horizon respectively) decreases for an increasing $\phi_i$ and approaches
zero in the limit $\phi_i\to 1/\sqrt{\xi_c}$.
\end{enumerate}
As seen in Figure 4, for regular solutions
$M$ approaches a finite value in the same limit.
In the black hole case, the critical value of $M$ increases very rapidly
with $r_h$,
which makes its accurate determination a difficult task for large $r_h$.
In the $k=-1$ case, the condition $H'(r_h)>0$ implies the existence of
a minimal value of the event horizon radius,
\begin{equation}
r_h>
\frac {(n-1)(n-3)(1-\xi\phi_h^2)}{[1+n(n-2)(1-\phi_h^2/8)]^{1/2} } .
\end{equation}

We have also found that non-trivial configurations may exist in the presence of a
nonzero scalar potential.
In this case the scalar field equation (\ref{KG}) in a fixed AdS background
has two exact solutions
\begin{eqnarray}
\label{exact-solutions}
\phi= \left[ 1+\frac{r^2}{\ell^2} \right]^p,~~V(\phi)=c\phi^s,
\end{eqnarray}
with $p=(2-n)/4,~c=-(n-2)^{3}/8n\ell ^{2},~s=2n/(n-2)$ in one case, and
$p=-n/4,~c=-n^3/8{\ell ^{2}}(n+2),~s=2(n+2)/n$ in the other.

However, we have restricted our analysis to the particular form
$V(\phi)=1/2\mu^2 \phi^n$, which for $n=4$ corresponds to the
case considered in \cite{Martinez:2004nb}.
These solutions share many properties with the zero-potential case,
being also indexed by the initial value of the scalar field $\phi_i$.
The  shape of the solutions is similar to the
$\mu=0$ case and we again found
no nodes in the scalar function.
In  Figure 1 we plotted a typical $n=4$ regular solution with
$\mu=0.3$. Similar solutions exist also in the black hole case.
In this context, we have found that
the black hole solution found in \cite{Martinez:2004nb}
corresponds to a $n=4,~k=-1$ configuration with a particular choice of
$(\mu,~\phi_h)$.

As expected, the mass $M$ of the solutions increases with $\mu$ while the
maximal value of $\phi_i$ decreases.
We will not address here the question of the
limiting solution for $\mu \neq 0$,
which seems to be an involved problem and
a different metric parametrization appears to be necessary.
Our preliminary results indicate
that for $\mu \neq 0$ the metric function $\delta(r_i)$
remains finite in this limit, while the value of $M$ diverges.

%%%%%%%%%%%%%%%%%%%%%%%%%%%%%%%%%%%%%%%%%%%%%%%%%%%%%%%%%%%%%%%
%% STABILITY
%%%%%%%%%%%%%%%%%%%%%%%%%%%%%%%%%%%%%%%%%%%%%%%%%%%%%%%%%%%%%%%
\section{On the stability of solutions}
\label{sec:stab}
\subsection{Stability of the numerical solutions}
Following the standard method, we consider spherically symmetric, linear
perturbations of our equilibrium solutions, keeping the metric ansatz as in
(\ref{metric}), but now the functions $H$, $\delta $ and $\phi $ depend on $t$
as well as $r$.
The algebra is simplest if we work in the transformed frame (see section 2.3), where
we have a minimally coupled scalar field, and, once the perturbation equations have been derived,
transform back to the frame with a conformally coupled scalar field.
The metric perturbations can be eliminated to yield a single perturbation equation for
\begin{equation}
\Psi = r^{\frac {(n-2)}{2}} \left( 1- \xi _{c} \phi ^{2} \right) ^{-\frac {1}{2}} \delta \phi ,
\end{equation}
where $\delta \phi $ is the perturbation in the conformally coupled scalar field.
For periodic perturbations ($\delta \phi (t,r) = e^{i\sigma t} \delta \phi (r)$, etc), the perturbation
equation takes the standard Schr\"odinger form
\begin{equation}
\sigma ^{2} \Psi = -\frac {d^{2}}{dr_{*}^{2}} \Psi + {\cal {V}} \Psi ,
\label{perteq}
\end{equation}
where we have defined the ``tortoise'' co-ordinate $r_{*}$ by
\begin{equation}
\frac {dr_{*}}{dr} = \frac {1}{He^{\delta }} .
\label{tortoise}
\end{equation}
For regular solutions, $r_{*}$ has values in a finite interval $[0,r_{*1}]$ for some
$r_{*1}<\infty $, while for black holes, $r_{*} \in (-\infty ,0]$.
The perturbation potential ${\cal {V}}$ is given as follows (we write
the formula explicitly only for the vanishing self-interaction potential case, for simplicity):
\begin{eqnarray}
{\cal {V}} & = &
\frac {H e^{2\delta }}{r^{2}}
\left\{
\frac {k}{2} (n-2)(n-3) - \frac {{\cal {A}}^{2}H}{(n-2)^{2} \Omega ^{2}}
-\frac {\Lambda r^{2}}{\Omega } + \frac {2\Lambda \xi _{c}nr^{2}}{(n-2)^{2}}
\left[ \frac {2n}{\Omega } - (n+2) \right]
+\frac {4\Lambda \xi _{c} n \phi  \phi' r^{3}}{{\cal {A}}\Omega }
\right.
 \nonumber  \\
& &  \left.
- \frac {k}{2}(n-2)^{3}(n-3) r^{2} \phi'^{2}
+\frac {\Lambda (n-2)^{2} r^{4} \phi '^{2}}{{\cal {A}}^{2} \Omega }
\right\},
\label{pertpot}
\end{eqnarray}
where $\Omega $ is given in (\ref{Omega}) and ${\cal {A}}$ is given in (\ref{calA}).

As is often the case for AAdS solutions, care is needed in the use of boundary conditions
to ensure that there is a self-adjoint operator in the perturbation equation (\ref{perteq})
(see, for example, the discussion in \cite{Bjoraker:2000qd} for the Einstein-Yang-Mills case).
We consider black hole and soliton solutions separately in this regard.

Firstly, for black hole solutions, it is convenient to change the independent variable to
$y=-r_{*}$ so that $y \in [0, \infty )$.
In order to have a self-adjoint operator, we need to impose the boundary condition
$\Psi =0 $ at $y=0$, which corresponds to $r\rightarrow \infty $.
Secondly, for the regular soliton solutions, boundary conditions need to be imposed
at both $r_{*}=0$ and $r_{*}=r_{*1}$ as we are working on a finite interval.
Suitable boundary conditions are $\Psi =0$ at both the end-points, namely at the origin
and at infinity.
In both these cases, it is straightforward to check that these
boundary conditions are sufficient to
enable a self-adjoint operator to be constructed from the differential operator
in (\ref{perteq}) using the standard techniques, found, for example, in section
XIII:2 of \cite{dunford}.

Some typical perturbation potentials (\ref{pertpot}) are shown in Figures 7--10.
We find a complicated behaviour of the perturbation potential depending on the values of the parameters
$\Lambda $, $\phi _{0}$ and $\mu $ and
the number of space-time dimensions.
As $r\rightarrow 0$, we have
\begin{equation}
{\cal {V}} \sim \frac {(n-4)(n-1)}{2r^{2}} + O(1),
\end{equation}
so for $n>4$ the potential diverges to infinity, like a standard central well potential with angular momentum
(see Figure 8).
The potential ${\cal {V}}$ can also be seen to vanish at the black hole event horizon (provided
${\cal {A}}>0$ and $\Omega >0$ there), and, at infinity, the
leading order behaviour is
\begin{equation}
{\cal {V}} \sim \frac {r^{2} n(n-4)}{4\ell ^{4}} +O(1),
\end{equation}
so for $n>4$ the potential again diverges to infinity (see Figures 8 and 10).
Turning on the self-interaction potential $V(\phi )$ tends to increase the perturbation potential (Figure 7),
as also observed in \cite{Winstanley:2002jt}.

In a limited number of cases we find that the perturbation potential (\ref{pertpot}) is positive everywhere
(see, for example, some of the plots in Figure 10).
In these cases, we can immediately conclude that the corresponding solutions are
(linearly) stable, since we have a self-adjoint operator in (\ref{perteq}).
However, for the majority of the solutions examined, the potential is not positive everywhere, and
in these cases we examine the {\em {zero mode}} solution of the perturbation equation (\ref{perteq}),
namely the time-independent solution when $\sigma ^{2}=0$
(see \cite{Winstanley:2005fu} for further details of the zero mode method applied to
scalar field perturbations).
For all the solutions we examined, the zero modes have no nodes (zeros).
As the operator in (\ref{perteq}) is self-adjoint, we can apply standard theorems
(see, for example, section XIII:7 of \cite{dunford}, or \cite{courant}),
which state that the number of nodes of the zero mode equals the number of
eigenvalues $\sigma^{2}$ of (\ref{perteq}) which
are less than zero.
Therefore there are no eigenmodes which grow exponentially in time and
the solutions are linearly stable.
Our conclusion that our numerical solutions are stable is in accord with previous work
\cite{Winstanley:2005fu,Torii:2001pg,Winstanley:2002jt}, in which solutions with a scalar field having no zeros
were linearly stable.

%%%%%%%%%%%%%%%%%%%%%%%%%%%%%%%%%%%%%%%%%%%%%%%%%%%%%%%%%%%%%%%%%%%%%%%%%
\subsection{Stability of the closed-form solutions}
We can also study the stability of the exact, closed form solutions on pure AdS space, given by
(\ref{exact-solutions}).
In this case we keep the background AdS metric fixed and perturb simply the scalar field.
The equation for the ``tortoise'' co-ordinate $r_{*}$ (\ref{tortoise})
can be explicitly integrated to give $x=r_{*}/\ell = \tan ^{-1} \left( r/\ell \right) $,
where $x\in [0,\pi/2]$, and the perturbation equation takes the form (\ref{perteq}),
for periodic perturbations, with $r_{*}$ replaced by $x$, and the perturbation potential is now
\begin{equation}
{\cal {V}} =
- \frac {1}{4} \left( n-2 \right) ^{2} \csc ^{2} (x)
- \frac {1}{2} \left( n-2 \right) \left( n-3 \right) \sec ^{2} (x)
-4p \left( p-1 \right)  ;
\end{equation}
where $4p(p-1)=(n-2)(n+2)/4$ for the first type of solutions, and
$4p(p-1)=n(n+4)/4$ for the second type of solutions.
Since the perturbation potential is so simple in this case, we use an analytic, variational, method
to study the stability.
We define a functional
\begin{eqnarray}
{\cal {F}} [\Psi ] & = & \int _{x=0}^{\frac {\pi }{2}} \left[  -\Psi \frac {d^{2}\Psi }{dx^{2}}
+ {\cal {V}} \Psi ^{2} \right] \, dx
\nonumber \\
& = &
\left[ - \Psi \frac {d\Psi }{dx} \right] _{x=0}^{\frac {\pi }{2}} +
\int _{x=0}^{\frac {\pi }{2}} \left[ \left( \frac {d\Psi }{dx} \right) ^{2}
+ {\cal {V}} \Psi ^{2} \right] \, dx ,
\label{variation}
\end{eqnarray}
where in the second line we have integrated by parts.
If we can find a test function $\Psi _{0}$ such that
\begin{equation}
\int _{x=0}^{\frac {\pi }{2}} \Psi _{0}^{2} \, dx
\label{integral}
\end{equation}
is finite, the boundary term in (\ref{variation}) vanishes and
${\cal {F}} [\Psi _{0}]<0$, then there must be at least one bound state
solution of the perturbation equation (\ref{perteq}) with $\sigma ^{2}<0$,
rendering the solutions unstable.
Using $\Psi _{0}(x) = \sin (2x) $ as our test function, the integral (\ref{integral}) is finite,
the required boundary term does indeed vanish, and
\begin{equation}
{\cal {F}} [\Psi _{0}] =
\left\{
\begin{array}{l}
\frac {1}{16} \left[ -13 n^{2} + 56 n - 44 \right]  ;
\\
\frac {1}{16} \left[ -13 n^{2} + 52 n - 48 \right]  ;
\end{array}
\right.
\end{equation}
for the first and second type of exact solutions, respectively.
Both the quadratics above are negative for all $n\ge 4$, so our exact,
 closed form solutions are unstable.
This instability is perhaps a little surprising given the stability of
the numerical solutions, however this
may be understood as a result of the particular form of the (negative--) scalar field potential.
If we perturb the scalar field slightly, there is no nearby solution
of the form (\ref{exact-solutions}) for
it to become.
The nearest numerical solution will require a finite change in the
scalar field at some value of $r$, and this
shows up as a linear instability.
However, it might be reasonable to expect that the scalar field
would settle on a stable, non-trivial
solution rather than radiating away to infinity.

%%%%%%%%%%%%%%%%%%%%%%%%%%%%%%%%%%%%%%%%%%%%%%%%%%%%%%%%%%%%%%%
%% CONCLUSIONS
%%%%%%%%%%%%%%%%%%%%%%%%%%%%%%%%%%%%%%%%%%%%%%%%%%%%%%%%%%%%%%%
\section{Conclusions}
\label{sec:conc}
In this paper we have studied the Einstein-scalar system in various
space-time dimensions, with a conformally coupled scalar field and
a negative cosmological constant.
We find both regular and black hole solutions, generalizing the black hole
solutions of \cite{Winstanley:2002jt}.

Both types of solutions are shown to be linearly stable,
apart from some exact, discrete, closed-form solutions on pure AdS, which are linearly unstable.
As with previous solutions \cite{Winstanley:2005fu,Winstanley:2002jt},
this stability can be readily understood in terms of the
Breitenlohner-Freedman bound \cite{Breitenlohner:1982jf}, which, in $n$ dimensions,
states that scalar fields in pure AdS are stable if their mass-squared
satisfies the inequality \cite{Breitenlohner:1982jf}
\begin{equation}
m^{2}_{BF}>\frac {\Lambda (n-1)}{2(n-2)},
\end{equation}
noting that $\Lambda <0$ so the bound is for negative mass-squared.
In our case, with zero self-interaction potential, the ``effective'' mass is given by
\begin{equation}
\xi _{c} R(r\rightarrow \infty ) = \frac {\Lambda n}{2(n-1)}.
\end{equation}
For all $n\ge 3$,  it is the case that $\xi _{c}R(r\rightarrow \infty )> m^{2}_{BF}$,
implying the stability of our solutions.

There are various interesting applications of these solutions.
The soliton solutions may be of interest in the gravitational collapse of scalar fields
in AAdS.
Critical collapse of a conformally coupled scalar field has been studied in flat space
\cite{choptuik}, but not, to date at least, in 3+1 dimensions in AAdS.
Soliton solutions of the type found in this paper do not occur in flat space, so
their presence in AAdS may change the phenomenology of gravitational collapse, since
there are no longer just the end-point possibilities of empty space or a black hole.
However, since the solitons we have found here are stable, they cannot be the
critical solutions, unlike the situation for Einstein-Yang-Mills solitons
in asymptotically flat space \cite{Choptuik:1996yg}.

It would also be of interest
to calculate the mass and action of these solutions, and to look for  possible
applications within the context of the AdS/CFT correspondence.
The thermodynamics of black holes with a conformally coupled scalar field with
a positive or zero cosmological constant has
already yielded some surprises (see, for example, \cite{Winstanley:2004ay} for a review),
prompting a detailed study of the thermodynamics when the cosmological constant is negative.

Finally, it would be interesting to extend the results derived in  \cite{Ashtekar:2003jh}
within the isolated horizon formalism, to the more general case considered
in this work.

We hope to return to these questions in a future publication.

\section*{Acknowledgements}
ER thanks Y Brihaye for collaboration during the early stages of this work
and DH Tchrakian for useful discussions.
EW would like to thank the Institute for Particle Physics Phenomenology, University of Durham,
for hospitality while this work was completed.
The work of ER was carried out in the framework of Enterprise--Ireland
Basic Science Research Project SC/2003/390.
The work of EW is supported by UK PPARC, grant reference number PPA/G/S/2003/00082.

%%%%%%%%%%%%%%%%%%%%%%%%%%%%%%%%%%%%%%%%%%%%%%%%%%%%%%%%%%%%%%%%%%%%%%%%%%%%%%

%%%%%%%%%%%%%%%%%%%%%%%%%%%%%%%%%%%%%%%%%%%%%%%%%%%%%%%%%%%%%%%%%%%%%%%%%%%%%%

%%%%%%%%%%%%%%%%%%%%%%%%%%%%%%%%%%%%%%%%%%%%%%%%%%%%%
\newpage
\setlength{\unitlength}{1cm}

\begin{picture}(15,15)
\centering
\put(-2,0){\epsfig{file=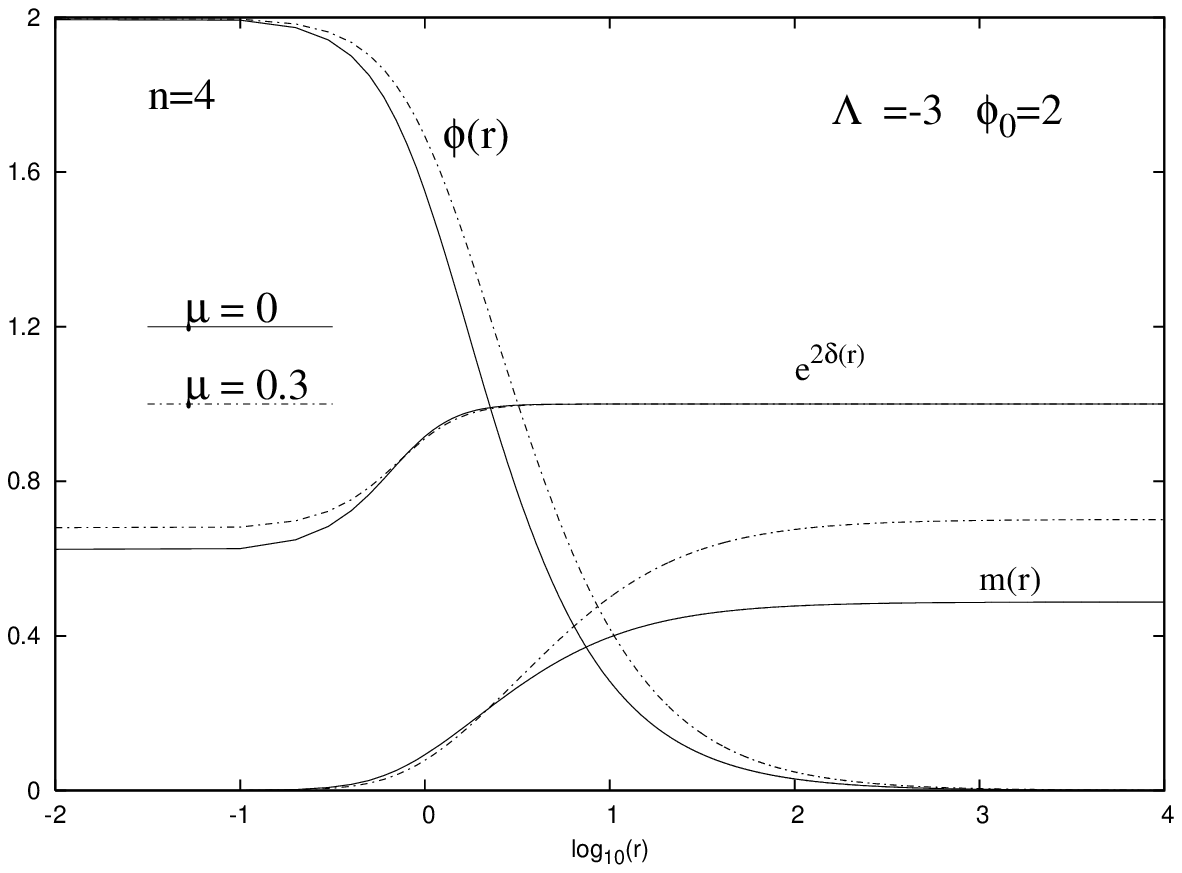,width=16cm}}
\end{picture}
\\
{\small {\bf Figure 1.}}
The functions $m(r),~e^{2\delta(r)}$ and $\phi(r)$
are plotted as functions of radius
for two typical $n=4$ regular  solutions.
The scalar potential in one case is $V(\phi)=\mu/2\phi^4$, with
$\mu=0.3$, while $V(\phi)=0$ for the second solution.
The central value of the scalar field is $\phi_0=2$.

%%%%%%%%%%%%%%%%%%%%%%%%%%%%%%%%%%%%%%%%%%%%%%%%%%%%%
\newpage
\setlength{\unitlength}{1cm}

\begin{picture}(15,15)
\centering
\put(-2,0){\epsfig{file=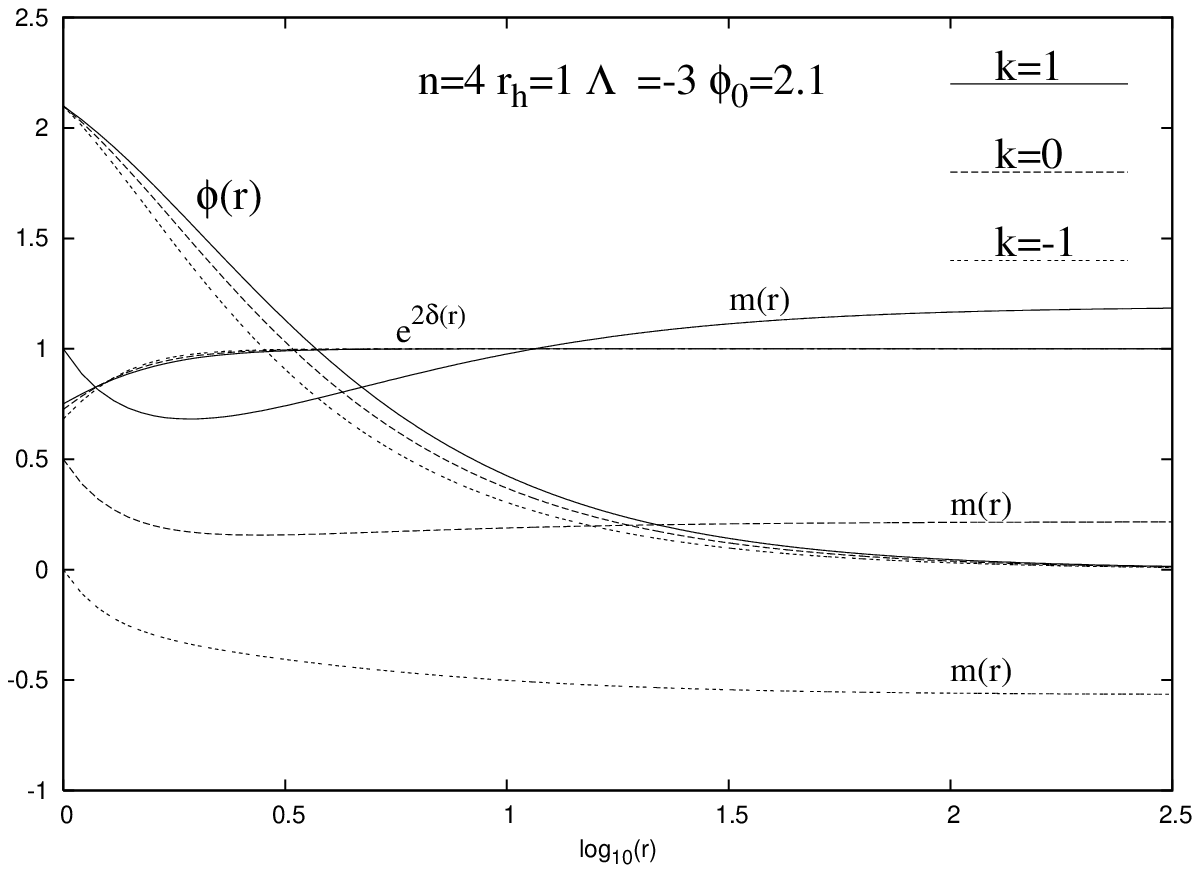,width=16cm}}
\end{picture}
\\
{\small {\bf Figure 2.}}
The functions $m(r),~e^{2\delta(r)}$ and $\phi(r)$
are plotted as functions of radius
for $n=4$ black hole solutions with $\phi_0=2.1$ and $V(\phi)=0$.

%%%%%%%%%%%%%%%%%%%%%%%%%%%%%%%%%%%%%%%%%%%%%%%%%%%%%
\newpage
\setlength{\unitlength}{1cm}

\begin{picture}(15,15)
\centering
\put(-2,0){\epsfig{file=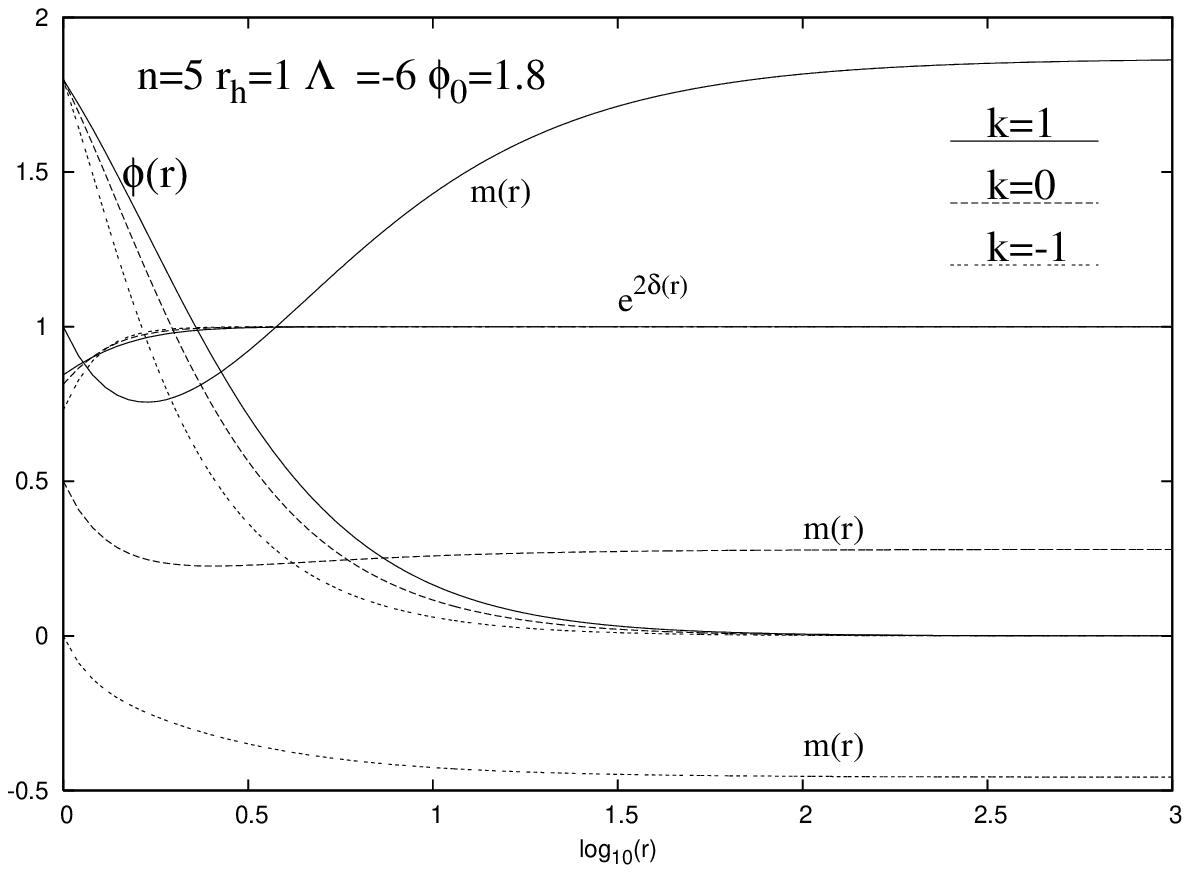,width=16cm}}
\end{picture}
\\
{\small {\bf Figure 3.}}
The functions $m(r),~e^{2\delta(r)}$ and $\phi(r)$
are plotted as functions of radius
for $n=5$ black hole solutions with $\phi_0=1.8$ and $V(\phi)=0$.

%%%%%%%%%%%%%%%%%%%%%%%%%%%%%%%%%%%%%%%%%%%%%%%%%%%%%
\newpage
\setlength{\unitlength}{1cm}

\begin{picture}(15,15)
\centering
\put(-2,0){\epsfig{file=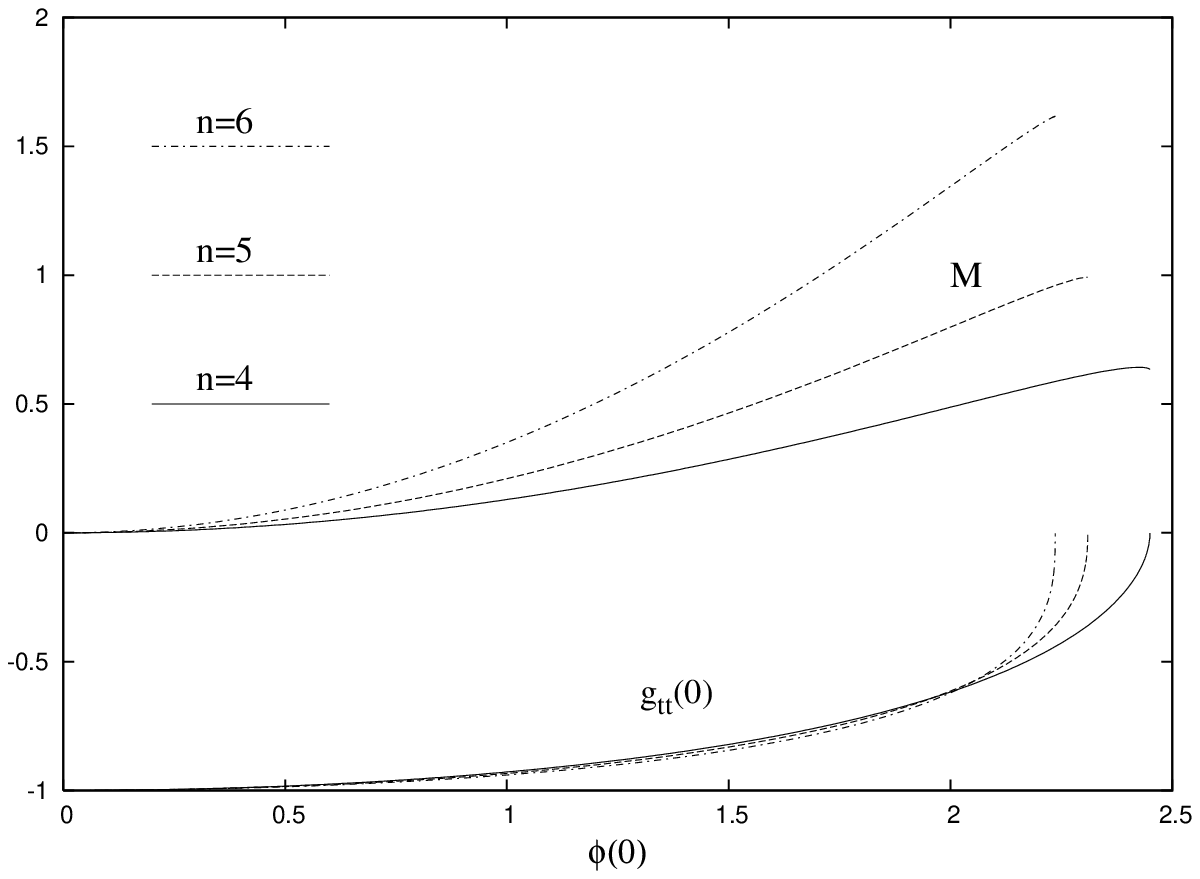,width=16cm}}
\end{picture}
\\
\\
{\small {\bf Figure 4.}}
The
parameters $M$ and $g_{tt}(0)=-e^{2\delta(0)}$
are shown as a function of $\phi(0)$ for
spherically symmetric regular solutions in several spacetime dimensions.
Here and in Figures 5, 6 the value of the cosmological constant is
$\Lambda=-(n-1)(n-2)/2$ and $V(\phi)=0$.

%%%%%%%%%%%%%%%%%%%%%%%%%%%%%%%%%%%%%%%%%%%%%%%%%%%%%
\newpage
\setlength{\unitlength}{1cm}

\begin{picture}(15,15)
\centering
\put(-2,0){\epsfig{file=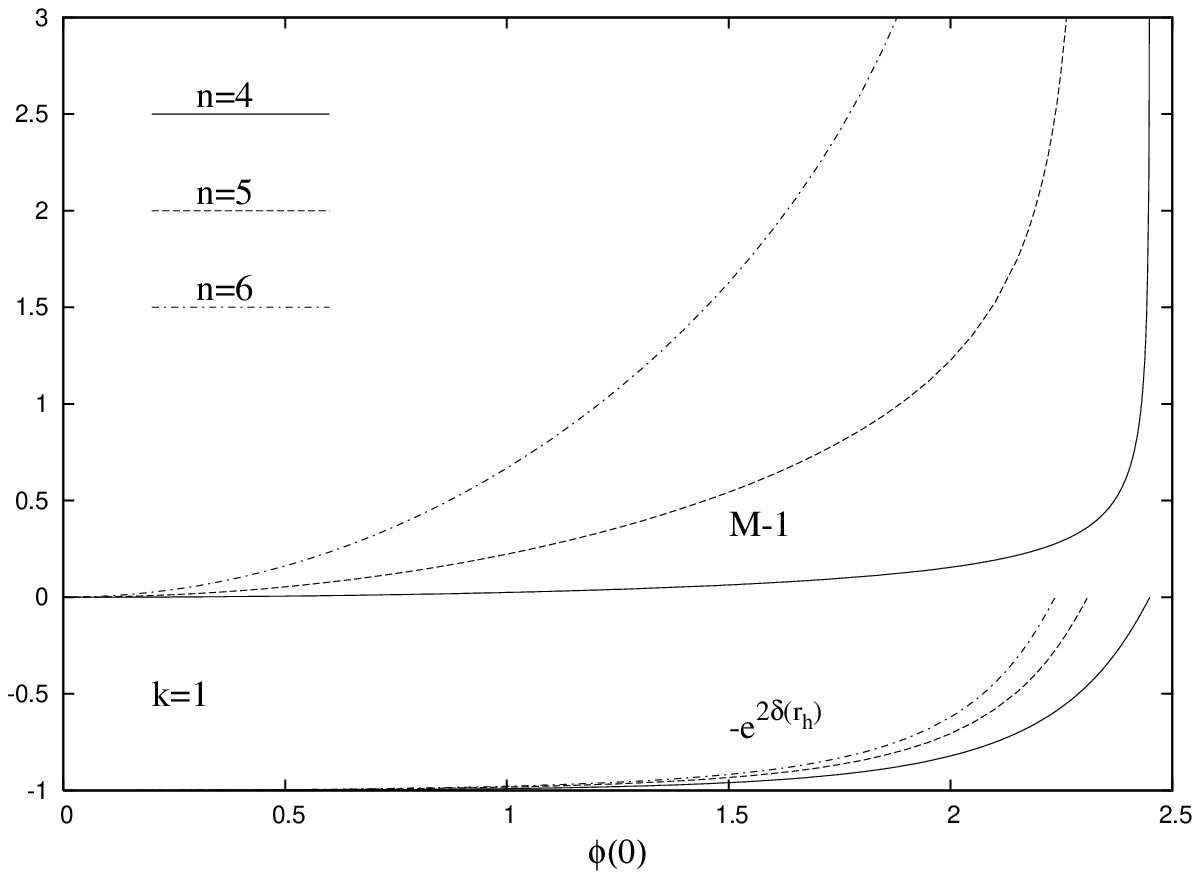,width=16cm}}
\end{picture}
\\
\\
{\small {\bf Figure 5.}}
The
parameters $M$ and $e^{2\delta(r_h)}$
are shown as a function of $\phi(r_h)$ for
spherically symmetric black hole solutions in several spacetime dimensions.

%%%%%%%%%%%%%%%%%%%%%%%%%%%%%%%%%%%%%%%%%%%%%%%%%%%%%
\newpage
\setlength{\unitlength}{1cm}

\begin{picture}(15,15)
\centering
\put(-2,0){\epsfig{file=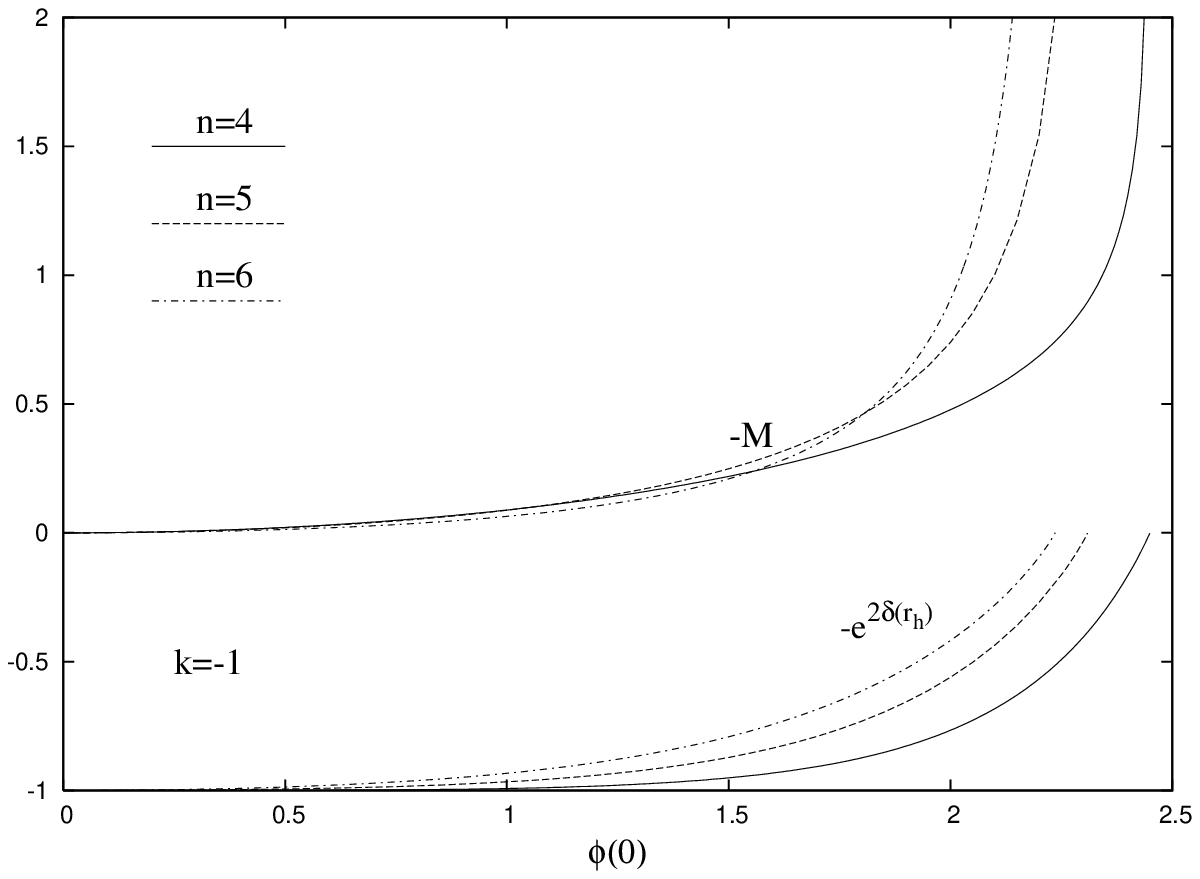,width=16cm}}
\end{picture}
\\
\\
{\small {\bf Figure 6.}}
The
parameters $M$ and $e^{2\delta(r_h)}$
are shown as a function of $\phi(r_h)$ for
$k=-1$ topological black hole solutions in several spacetime dimensions.

%%%%%%%%%%%%%%%%%%%%%%%%%%%%%%%%%%%%%%%%%%%%%%%%%%%%%
\newpage
\setlength{\unitlength}{1cm}

\begin{picture}(15,15)
\centering
\put(-2,12){\epsfig{file=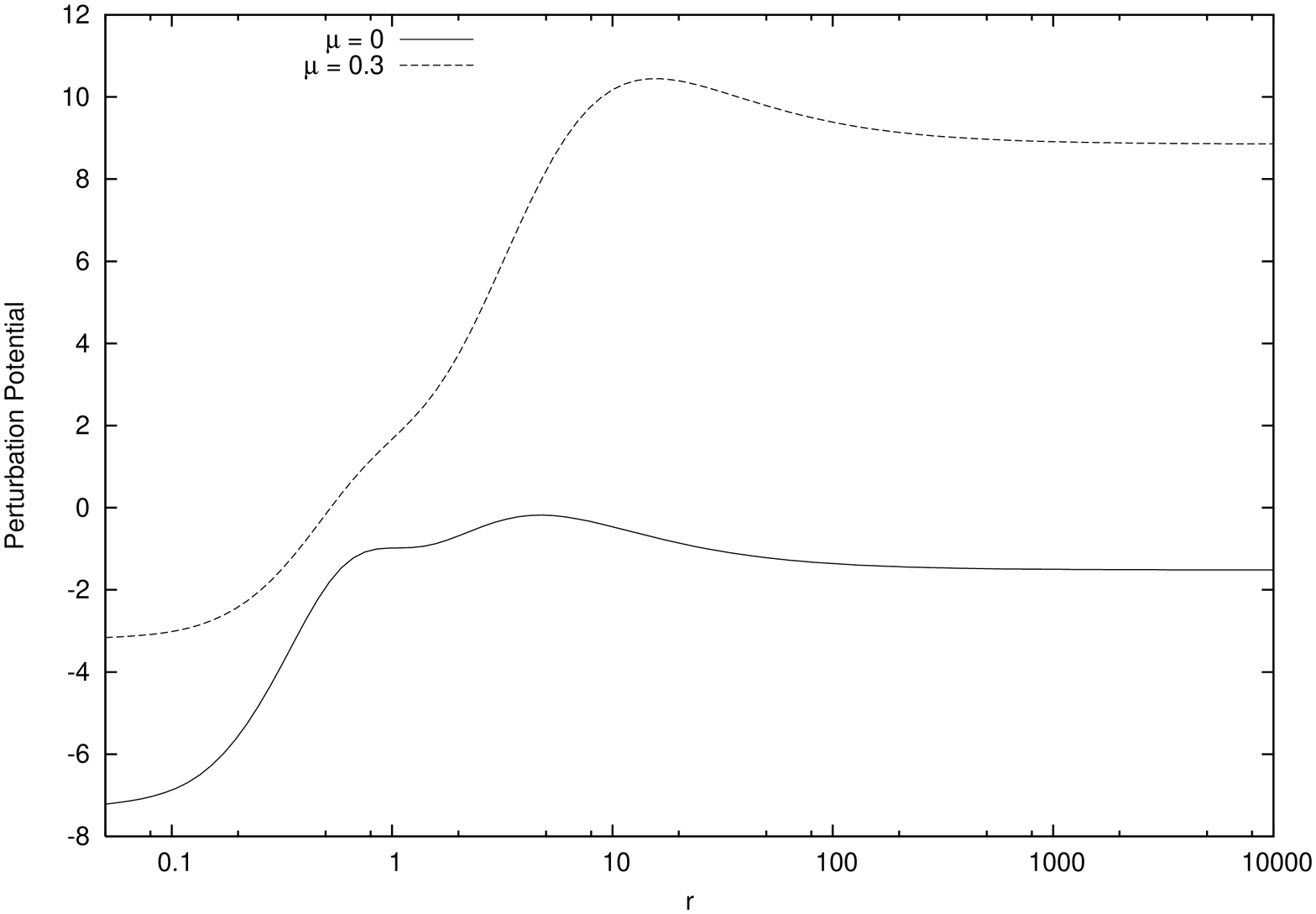,angle=270,width=16cm}}
\end{picture}
\\
\\
{\small {\bf Figure 7.}}
The perturbation potential (\ref{pertpot}) is plotted for
the typical $n=4$ regular solutions shown in Figure 1.

%%%%%%%%%%%%%%%%%%%%%%%%%%%%%%%%%%%%%%%%%%%%%%%%%%%%%
\newpage
\setlength{\unitlength}{1cm}

\begin{picture}(15,15)
\centering
\put(-2,12){\epsfig{file=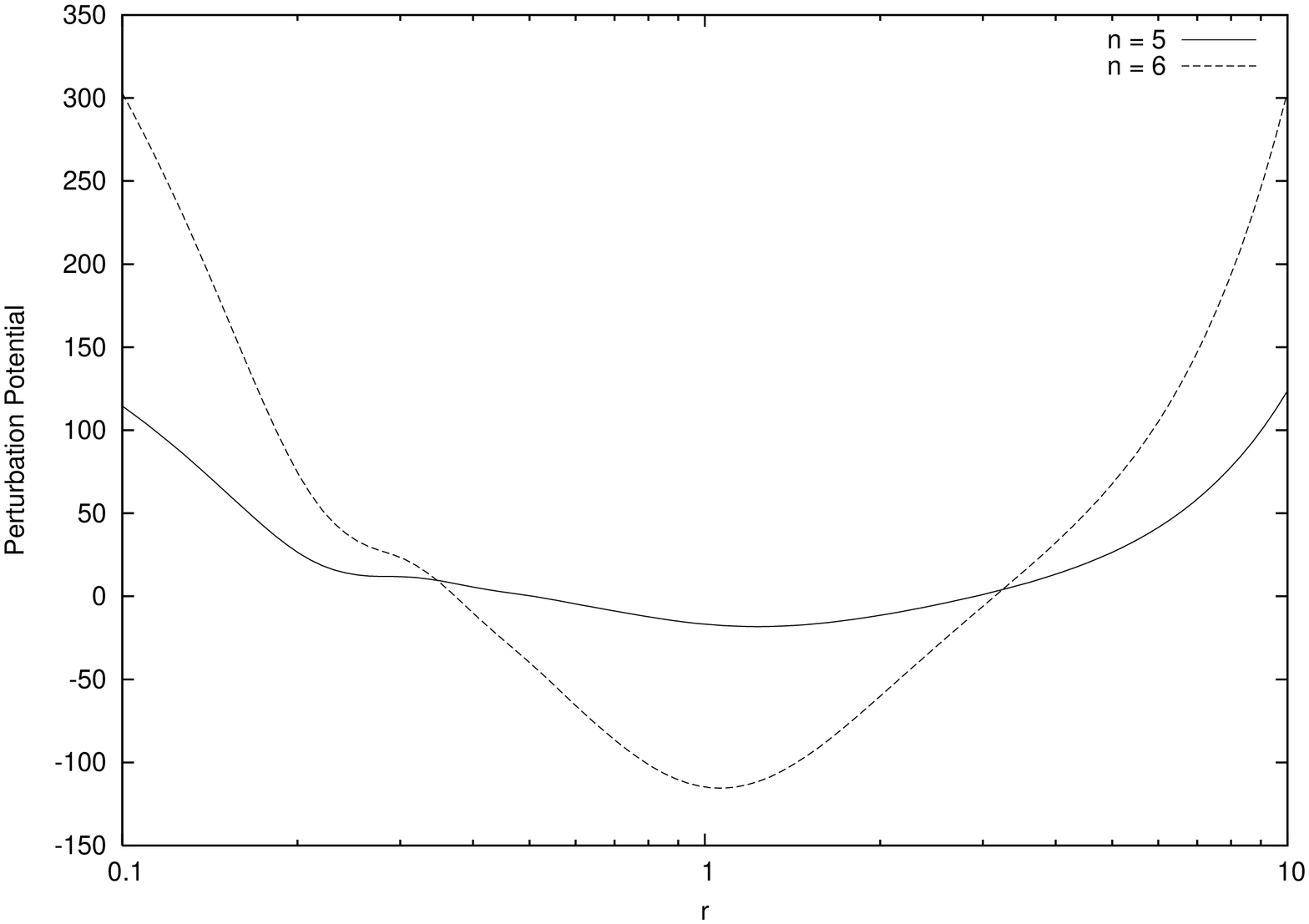,angle=270,width=16cm}}
\end{picture}
\\
\\
{\small {\bf Figure 8.}}
The perturbation potential (\ref{pertpot}) is plotted for
two typical $n=5,6$ regular solutions.
For both the solutions, the scalar potential $V(\phi )=0$, and the
central value of the scalar field is $\phi _{0}=2$.

%%%%%%%%%%%%%%%%%%%%%%%%%%%%%%%%%%%%%%%%%%%%%%%%%%%%%
\newpage
\setlength{\unitlength}{1cm}

\begin{picture}(15,15)
\centering
\put(-2,12){\epsfig{file=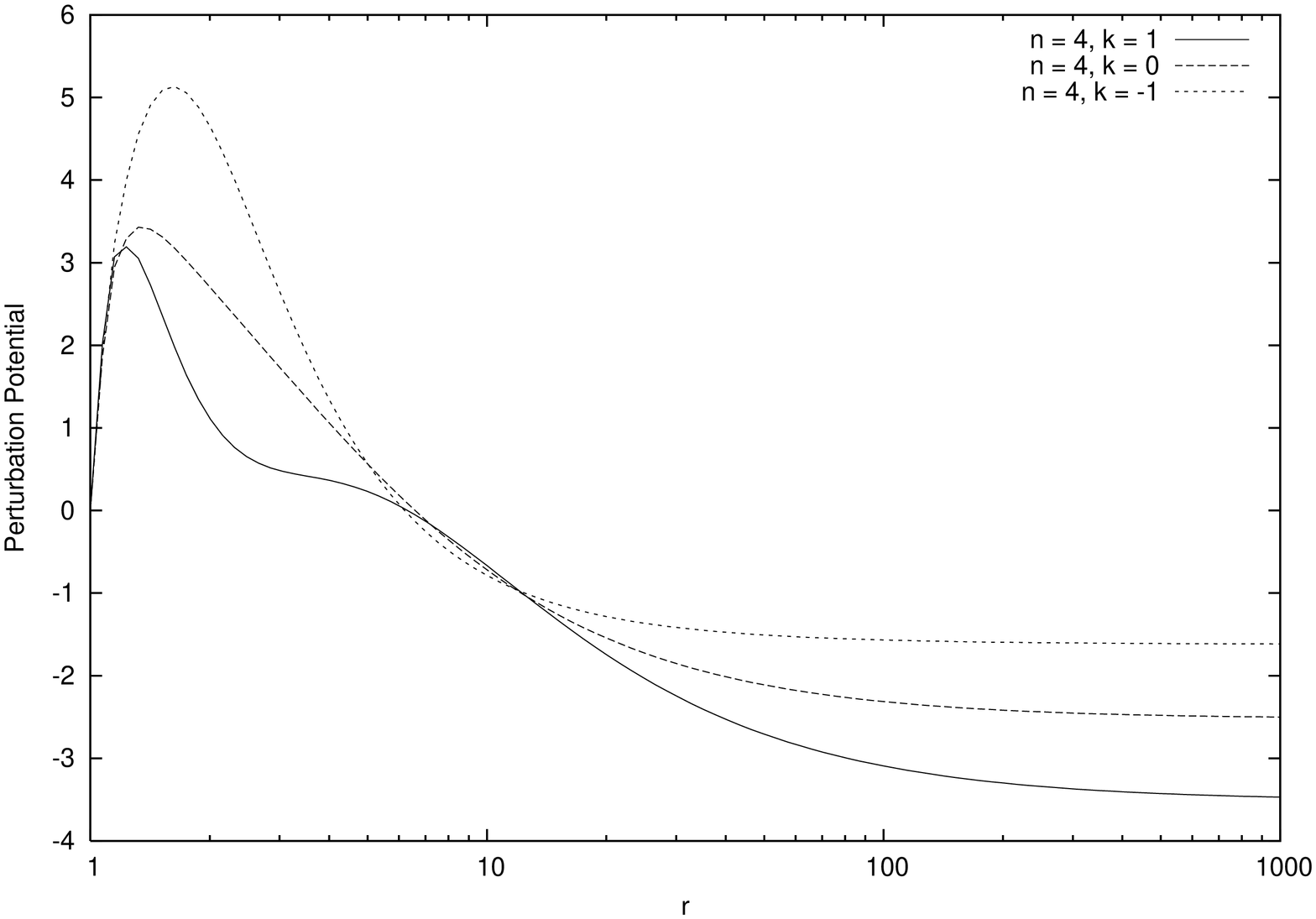,angle=270,width=16cm}}
\end{picture}
\\
\\
{\small {\bf Figure 9.}}
The perturbation potential (\ref{pertpot}) is plotted for
the typical $n=4$ black hole solutions shown in Figure 2.

%%%%%%%%%%%%%%%%%%%%%%%%%%%%%%%%%%%%%%%%%%%%%%%%%%%%%
\newpage
\setlength{\unitlength}{1cm}

\begin{picture}(15,15)
\centering
\put(-2,12){\epsfig{file=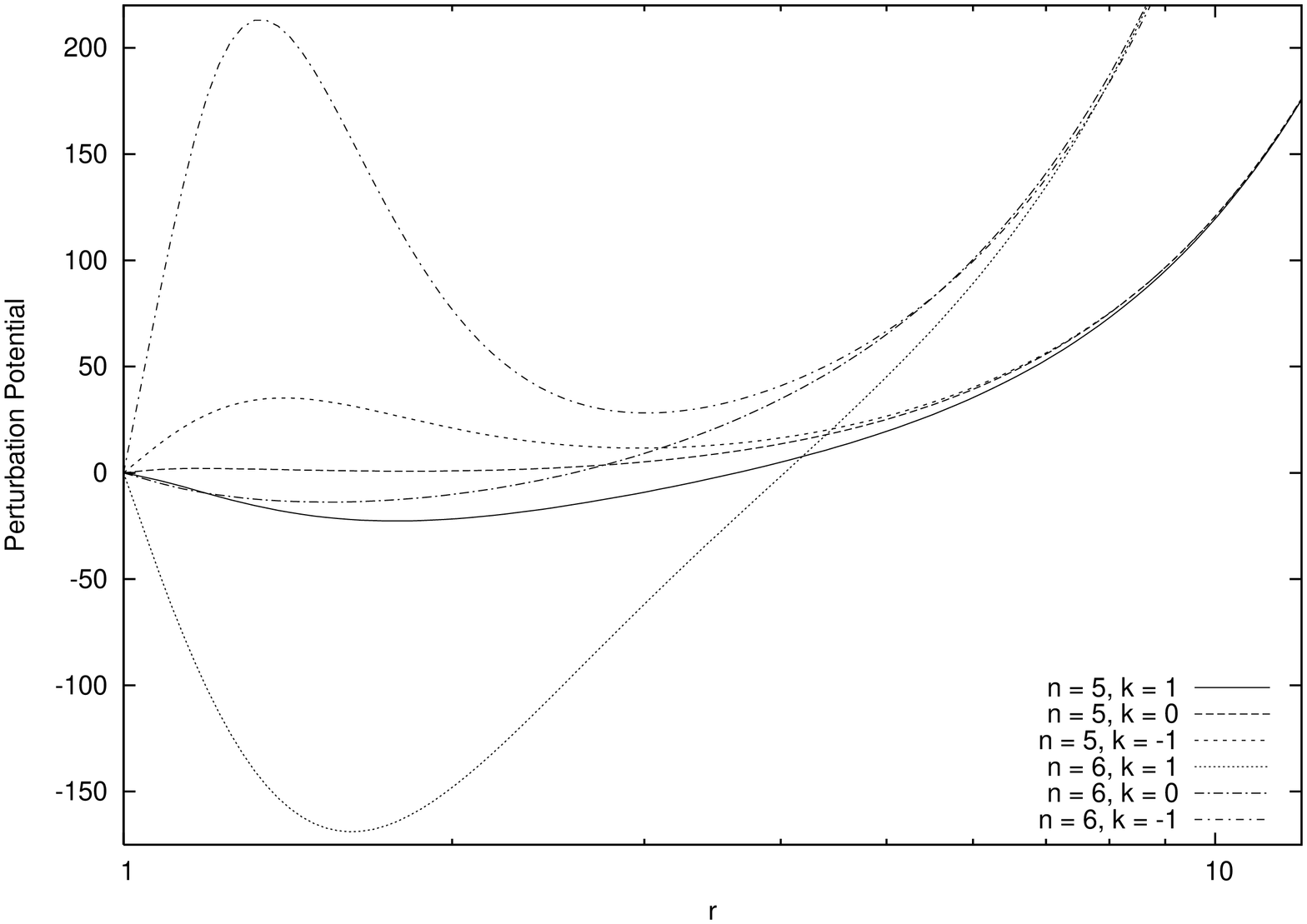,angle=270,width=16cm}}
\end{picture}
\\
\\
{\small {\bf Figure 10.}}
The perturbation potential (\ref{pertpot}) is plotted for
typical $n=5,6$ black hole solutions with the value
of the scalar field on the event horizon $\phi _{0}=1.8$.

\end{document}